\documentclass[namedreferences]{solarphysics}
\usepackage[hyperref,optionalrh,solaromanenum]{spr-sola-addons} 
\usepackage{graphicx}                    
\usepackage{color}                       
\usepackage{breakurl}                         

\begin{document}

\begin{article}
\begin{opening}
\title{Centre to limb brightness variations from ALMA full disk solar images}
\author[addressref=aff1,corref,email={davor.sudar@gmail.com}]{\inits{D.}\fnm{Davor}~\lnm{Sudar}~\orcid{0000-0002-1196-6340}}
\author[addressref=aff1,email={romanb@geof.hr}]{\inits{R.}\fnm{Roman}~\lnm{Braj\v{s}a}}
\author[addressref=aff1,email={ivica.skokic@gmail.com}]{\inits{I.}\fnm{Ivica}~\lnm{Skoki\'{c}}}
\author[addressref={aff2,aff3},email={benz@astro.phys.ethz.ch}]{\inits{A.O.}\fnm{Arnold O.}~\lnm{Benz}}

\address[id=aff1]{Hvar Observatory, Faculty of Geodesy, Ka\v{c}i\'{c}eva 26, University of Zagreb, 10000 Zagreb, Croatia}
\address[id=aff2]{University of Applied Sciences and Arts Northwestern Switzerland, Bahnhofstrasse 6, 5210 Windisch, Switzerland}
\address[id=aff3]{Institute for Particle Physics and Astrophysics, ETH Zurich, 8093 Zurich Switzerland}

\runningauthor{D. Sudar et al.}
\runningtitle{Centre to limb brightness variations}

\begin{abstract}
Science Verification (SV) data of solar observations with Atacama Large Millimeter-submillimeter Array (ALMA) telescope were
released to the scientific community. Understanding the centre to limb brightness function is necessary to compare features in full disk images.
Our goals are to find the empirical centre to limb brightness functions in two available spectral bands and create flattened images
with centre to limb brightness variations removed.
We used second-order polynomial fit of the cosine of incidence angle to data points as a function of radial distance to the centre of
the solar disk. The method also includes iterative removal of outliers based on the interquartile range.
Fitting functions for all available images proved to adequately describe the data with comparatively small errors in the fitting coefficients.
In both bands we found brightening towards the limb which is a consequence
of increase in electron temperatures with radial distance
in this region of the solar atmosphere.
This study found that the $T_{b}$ of an active region has about 180 K difference between with and without the limb brightening
at radial distance $\approx$0.75$R_{\odot}$ in Band 6.
We also made flattened images with limb brightening
removed.
The limb brightening effect in ALMA images is significant enough (of the order of 10\% for Band 3 and about 15\% in Band 6)
that it can not be neglected in further analyses. Since the effect of the side lobes was not included in this study, these
values probably represent the lower limit of the limb brightening. The shape of the limb brightening function can also be used to constrain electron
densities and temperatures in various layers of the solar atmosphere.
\end{abstract}

%

\end{opening}

%
\section{Introduction}
The Atacama Large Millimeter-submillimeter Array (ALMA) telescope is currently capable of observing the Sun in three
spectral bands: B3 ($\approx$100 GHz), B6 ($\approx$250 GHz), and B7 ($\approx$300 GHz). The array of antennas can make interferometric images of small areas with
high spatial resolution \citep{Shimojo2017}, but also create a full disk images with a lower resolution \citep{White2017}.
Full disk images are regularly supplied with each observation of the Sun.

In the visual part of the solar spectrum, brightness variations towards the limb are known as limb darkening.
In the ALMA wavelength range we expect to see brightening towards the solar limb. The explanation of this effect lies in the fact
that photons 'escape' from solar atmosphere where the optical depth, $\tau$, is $\approx$1.
When looking at the centre we observe deeper layers than on the limb due to the projection of the solar sphere on the sky plane.
If the temperature of the deeper layer is larger, which is the case for the visual part of the spectrum, the centre is brighter
than the limb and we observe limb darkening.
Somewhere in the chromosphere the temperature gradient must change since we know that the outermost layer of the Sun, the corona,
is hot ($>$1 million K). At wavelengths in the mm range we are observing the chromosphere and we expect that the
deeper layers, observed in the centre, are cooler than higher layers seen at the limb, so we should see limb brightening.
In fact limb brightening was already observed with ALMA \citep{White2017, Alissandrakis2017, Brajsa2017, Selhorst2019} and with other instruments observing
at similar wavelengths \citep{Bastian1993b, Brajsa1994, Lindsey1995, Nindos1999, Pohjolainen2000a}.

The observed brightness at ALMA wavelengths is usually given as brightness temperature, $T_{b}$, and images produced by ALMA
are given in Kelvins.
Comparing $T_{b}$ of different features versus the quiet Sun level even on a single image can prove to be problematic when centre to limb brightness
variations are not taken into account \citep{Brajsa2017}.
Measuring brightness variations over time of a single object \citep{Pohjolainen1991} would also create problems, because the position on the solar disk
changes. Comparing $T_{b}$ in different bands of the same object is also affected, unless the position is the same in both bands.
A number of features show fairly small contrast against the quiet Sun level in or near the ALMA wavelength range. Sometimes
they are seen as darker and sometimes as brighter features. Typical examples are observations of filaments on the disk \citep{Kundu1978, Schmahl1981,
Hiei1986, Bastian1993a, Brajsa2017},
coronal holes \citep{Kundu1976, Kosugi1986, Gopalswamy1999, Pohjolainen2000b, White2006, Brajsa2007, Brajsa2017} and magnetic inversion lines
\citep{Vrsnak1992, Bastian1993a, Brajsa2017}. Properly accounting for the centre to limb brightness variations might help to better
and more consistently detect these features.

Our goal in this work is to measure accurate brightness variations from the solar centre to its limb using full disk images
released as part of Science Verification (SV) observations \citep{Bastian2015, Kobelski2016} available at the ALMA science portal
(https://almascience.nrao.edu/alma-data/science-verification).
We will also provide an empirical function, in the form of the polynomial expansion of $\cos{\psi}$ \citep{Neckel1994}
fitting the average observed brightness as a function of distance from the centre for the
two bands and create so-called flattened images with limb brightening removed. These images are then suitable for various analyses
regardless of the object's position on the solar disk.
In previous studies, \citet{Alissandrakis2017} and \citet{Selhorst2019}
calculated the centre to limb brightness variations as a series of average values depending on the radial distance from the centre of the Sun.
In addition, \citet{Alissandrakis2017} masked some areas of the full disk image to remove features which they considered too bright or too dark.
However, neither \citet{Alissandrakis2017} nor \citet{Selhorst2019} attempted to remove the centre to limb brightness variations from the observed image.

\section{Data and Methods}

Our data set consists of six Band 3 and three Band 6 full disk images obtained during solar
SV campaign in the period from 16th to 20th December 2015 (Table 1).
The images were recorded with 12 m ALMA total power antennas using the single-dish fast-scanning method with a double circle pattern \citep{White2017}.
The complete scan of the Sun (including calibration) takes around 13 (17) minutes in Band 3 (6) with 1 ms sample time.
The data were calibrated in CASA using the sdcal task, and the final brightness temperature was corrected for the antenna efficiency \citep{White2017}.
Measurement samples were re-gridded onto the rectangular mesh with a cell size of 6$^{\prime\prime}$ and 3$^{\prime\prime}$ per pixel, for Bands 3 and 6, respectively.
Band 3 images were all made in spectral window no. 3 with a primary beam FWHM of 58$^{\prime\prime}$ at frequency $\nu=107$ GHz ($\lambda=2.8$ mm).
All Band 6 images were also made in spectral window no. 3 with a primary beam FWHM of 25$^{\prime\prime}$ at frequency $\nu=248$ GHz ($\lambda=1.2$ mm).
Images obtained with this procedure are identical to reference images provided by the ALMA team for each observation to within 1 K in each
pixel.

Since there are large variations in absolute calibrations between individual datasets, \citet{White2017} recommended
that the images are scaled to match their best determination of the quiet Sun temperatures in the disk centre. Similar to what \citet{White2017}
suggested we modified the images by taking the average $T_{b}$ in circle around the centre covering the same square area as given by \citet{White2017}
and rescaling them to match 5900 K and 7300 K in the centre in Bands 6 and 3, respectively.
\begin{table}
\caption{Time of observation, frequency and wavelength of the data.}              
\label{Tab_Data}      
\begin{tabular}{c c c c c l}          
\hline
ID & Band & Date and time [UT] & $\nu$ [GHz] & $\lambda$ [mm] & Execution block ID\\    
\hline                                   
  1 & 3 & 2015-12-16 18:27:42 & 107 & 2.8 & uid://A002/Xade68e/X40a\\
  2 & 3 & 2015-12-16 19:42:06 & 107 & 2.8 & uid://A002/Xade68e/X7c8\\
  3 & 3 & 2015-12-17 19:12:58 & 107 & 2.8 & uid://A002/Xade68e/X40f7\\
  4 & 3 & 2015-12-17 19:20:37 & 107 & 2.8 & uid://A002/Xade68e/X40fe\\
  5 & 3 & 2015-12-17 19:28:02 & 107 & 2.8 & uid://A002/Xade68e/X414d\\
  6 & 3 & 2015-12-17 19:35:27 & 107 & 2.8 & uid://A002/Xade68e/X4154\\
  7 & 6 & 2015-12-17 14:52:36 & 248 & 1.2 & uid://A002/Xade68e/X3ad4\\
  8 & 6 & 2015-12-18 20:12:21 & 248 & 1.2 & uid://A002/Xae00c5/X2e6b\\
  9 & 6 & 2015-12-20 13:52:40 & 248 & 1.2 & uid://A002/Xae17cd/X373c\\
\hline                                             
\end{tabular}
\end{table}

\begin{figure}
\resizebox{\hsize}{!}{\includegraphics{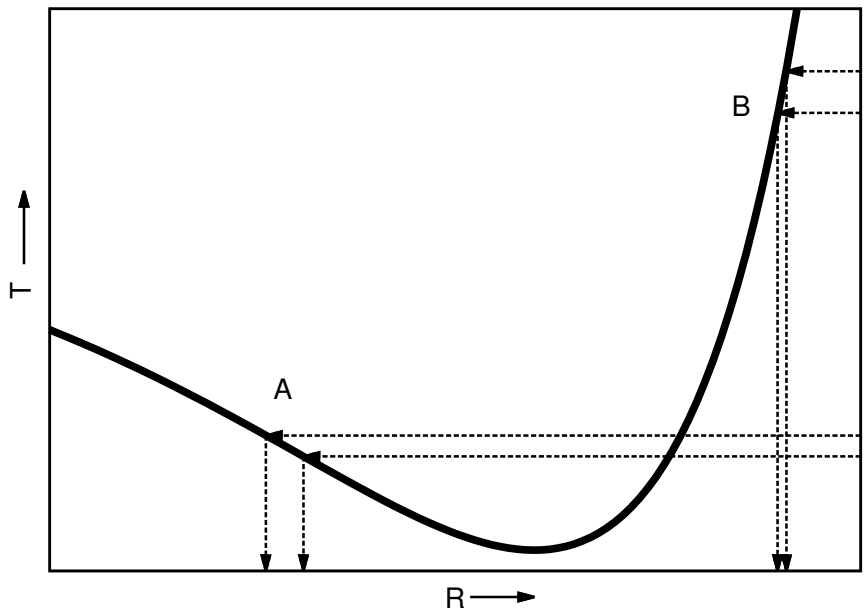}}
\caption{Schematic explanation of different types of centre to limb brightness variations based
on variation of temperature with height. The thick solid line represents temperature, $T$,
with height above the photosphere, $R$. In the case denoted with A, the temperature gradient is negative,
while in the case B, the gradient is positive. Horizontal arrows show the effective depth of observed layers,
where the arrow penetrating into deeper layers is in the centre of the solar disk in both cases.}
\label{Fig_TauScheme} 
\end{figure}
Data analysis in this paper only considers the main beam of the telescope. \citet{Iwai2017} discussed
the effect of the side lobes of the 45 m Nobeyama radio telescope on the observed centre to limb brightness profile.
While the ALMA 12 m antennas yield beam patterns with significantly smaller side lobes the effect is still there, so we should
bear in mind that the observed profile is not exactly the same as the true centre to limb brightness variations.

As mentioned in the previous section, the limb function is sensitive to the gradient of the temperature, $T$, with distance from the photosphere, $R$.
In Figure~\ref{Fig_TauScheme} we show a schematic explanation of observed limb brightness function depending on the temperature gradient.
The thick solid line represents a schematic view of temperature variations with radial distance from the surface of the Sun. Closer to the surface,
the temperature drops (left part of the plot), while further out the temperature increases (right part of the plot) in order to
reach millions of Kelvin in the corona. Photons emitted in the direction of the observer from the centre of the solar disk originate
in deeper layers than those originating closer to the solar limb because they have to pass through thinner layers of the solar atmosphere in the direction of the
observer.
The horizontal arrows in Figure~\ref{Fig_TauScheme} show which layer we observe in the centre and on the limb for two different cases of the temperature gradient. The arrow
penetrating into deeper layers is the one corresponding to the observed temperature at the centre in both cases.
In case A of the Figure~\ref{Fig_TauScheme} we see a negative gradient and as a consequence observe limb darkening (higher atmosphere layer is cooler
then the lower layer), 
while in case B part of the image, temperature gradient is positive and limb brightening is observed (higher atmosphere layer is hotter
than the lower layer).

\begin{figure}
\resizebox{\hsize}{!}{\includegraphics{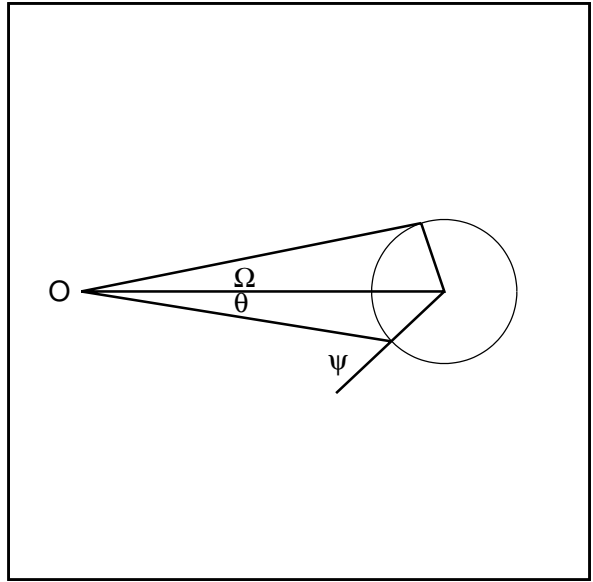}}
\caption{Geometry of the observer, O, and the Sun. The incidence
angle, $\psi$, is the angle between the observer's line of sight and the normal
to the solar surface. The observer sees this point at the angle $\theta$
from the centre of the solar disk. $\Omega$ is the angle at which
the observer sees the limb of the Sun.}
\label{Fig_Geometry} 
\end{figure}
The geometry of the system is shown in Figure~\ref{Fig_Geometry}.
The intensity at angle $\psi$ can be expressed as a polynomial in $\cos\psi$ (see for example \citet{Neckel1994}):
\begin{equation}
\frac{T_{b}(\psi)}{T_{b}(0)}=\sum_{k=0}^{N}a_{k}\cos^{k}{\psi},
\label{Eq_LimbDarkF}
\end{equation}
where $T_{b}(0)$ is the brightness temperature in the centre of the solar disk,
$k$ is the exponent, $a_{k}$ are the coefficients of the polynomial expansion, and $N$ is the order
of the polynomial.

Using the geometrical relationships from Figure~\ref{Fig_Geometry} we can express the incidence angle as a function
of the observed angle:
\begin{equation}
\cos\psi=\sqrt{1 - \left(\frac{\sin\theta}{\sin\Omega}\right)^{2}}.
\end{equation}
Substituting angles with distances from the centre of the solar disk we finally obtain:
\begin{equation}
\cos\psi \approx\sqrt{1 - \left(\frac{r}{R_{\odot}}\right)^{2}},
\label{Eq_CosRinPix}
\end{equation}
where the distance from the centre, $r$, and the solar radius $R_{\odot}$ are given in pixels.
We used only the terms up to the second order of $\cos\psi$ in Equation~(\ref{Eq_LimbDarkF}), so that
the function for fitting to the observed data is of the form:
\begin{equation}
T_{b}(\psi)=A_{0} + A_{1}\cos{\psi} + A_{2}\cos^{2}{\psi}.
\label{Eq_FitF}
\end{equation}

Using the angular solar diameter values obtained for the correct time frame from JPL
Horizons tool\footnote{\url{https://ssd.jpl.nasa.gov/horizons.cgi}}, we get a {\em photospheric}
solar radius of 162.5 and 325.0 pixels for Band 3 and Band 6, respectively.
The chromosphere, which we observe with ALMA, is about 2000 km thick. Therefore the radii 
are very close to their photospheric values (difference
is about half a pixel in Band 3 and about 1 pixel in Band 6 which is about 3$^{\prime\prime}$ in both bands).
The chromospheric radii, $R_{\odot}$, which we will be using
in the rest of the paper are 162.8 and 325.6 pixels in Band 3 and Band 6, respectively.

Full disk images of the Sun in ALMA bands 3 and 6 show many variations and details with
differences up to $\approx$1000 K \citep{Brajsa2017}. Therefore, we devised
a method which can remove pixels with large deviations from the mean value while simultaneously
fitting a function given in Equation~(\ref{Eq_FitF}). The method consists of iterative fitting and removing the outliers
based on the interquartile range which is described in more details in \citet{Sudar2016, Sudar2017} who used
this method in another context.
The iterative procedure stops when there are no more outliers to remove and in this case it took
about 10 iterations for each image in both bands. The main difference from the method described
in \citet{Sudar2016, Sudar2017} is that we have chosen $k=1.5$ which is more restrictive and removes
more data points.

\begin{figure}
\resizebox{\hsize}{!}{\includegraphics{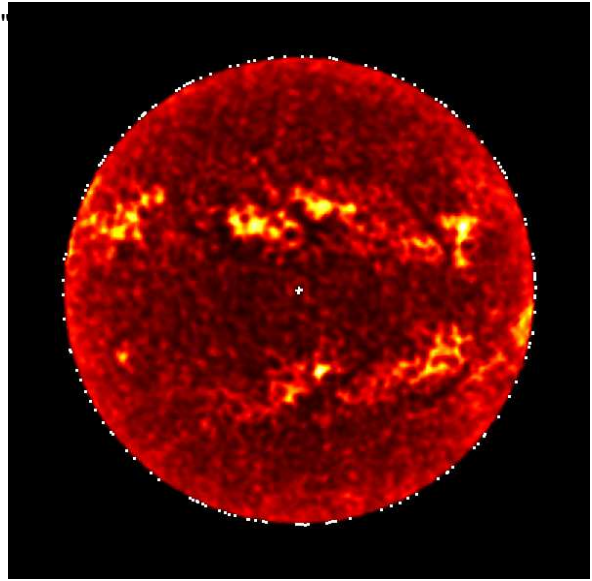}}
\caption{Superimposed on an image taken in Band 6 on Dec 18th 2015 (ID = 8, in Table \ref{Tab_Data})
we show small white solid squares near the solar limb which represent all data points
with a brightness temperature, $T_{b}$, between 70-71\% of the maximum value in the image.
The white cross is the centre of the solar disk obtained with the steepest descent method.}
\label{Fig_FindCenter}
\end{figure}
Important part of this analysis is to find the accurate position of the Sun's centre in each image.
First we select pixels in the range where the $T_{b}$ is between 70\% and 71\% of the maximum $T_{b}$.
They lie on a circle around the centre (Figure~\ref{Fig_FindCenter}). Actual level values are not
very important as long as they make a circle far away from the centre. In the next step we use
the steepest descent algorithm to optimize the value of the centre position by minimising the function
$\sum(R_{i} - \overline{R})^{2}$, where $R_{i}$ is the distance of the pixel on the circle to the trial
central position and $\overline{R}$ is the average value for all pixels on the circle.

\section{Results}

\begin{figure}
\resizebox{\hsize}{!}{\includegraphics{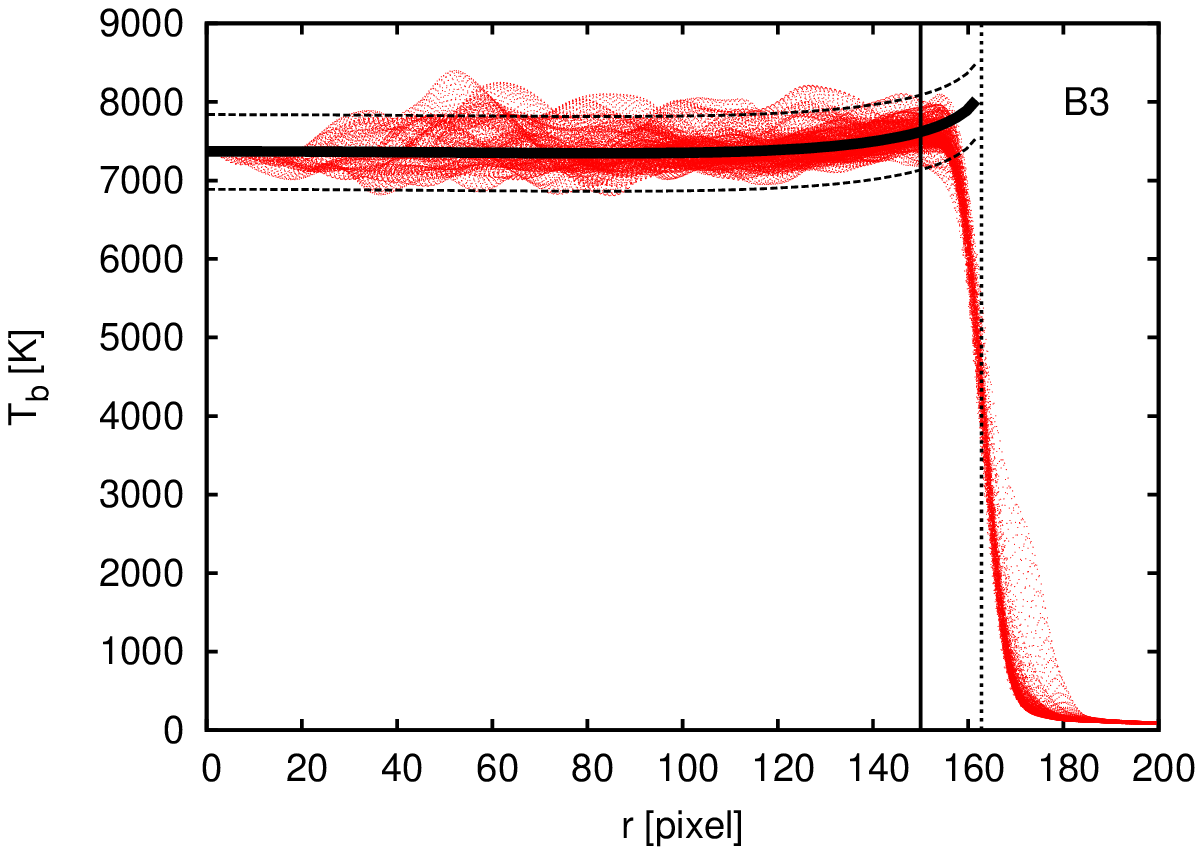}}
\resizebox{\hsize}{!}{\includegraphics{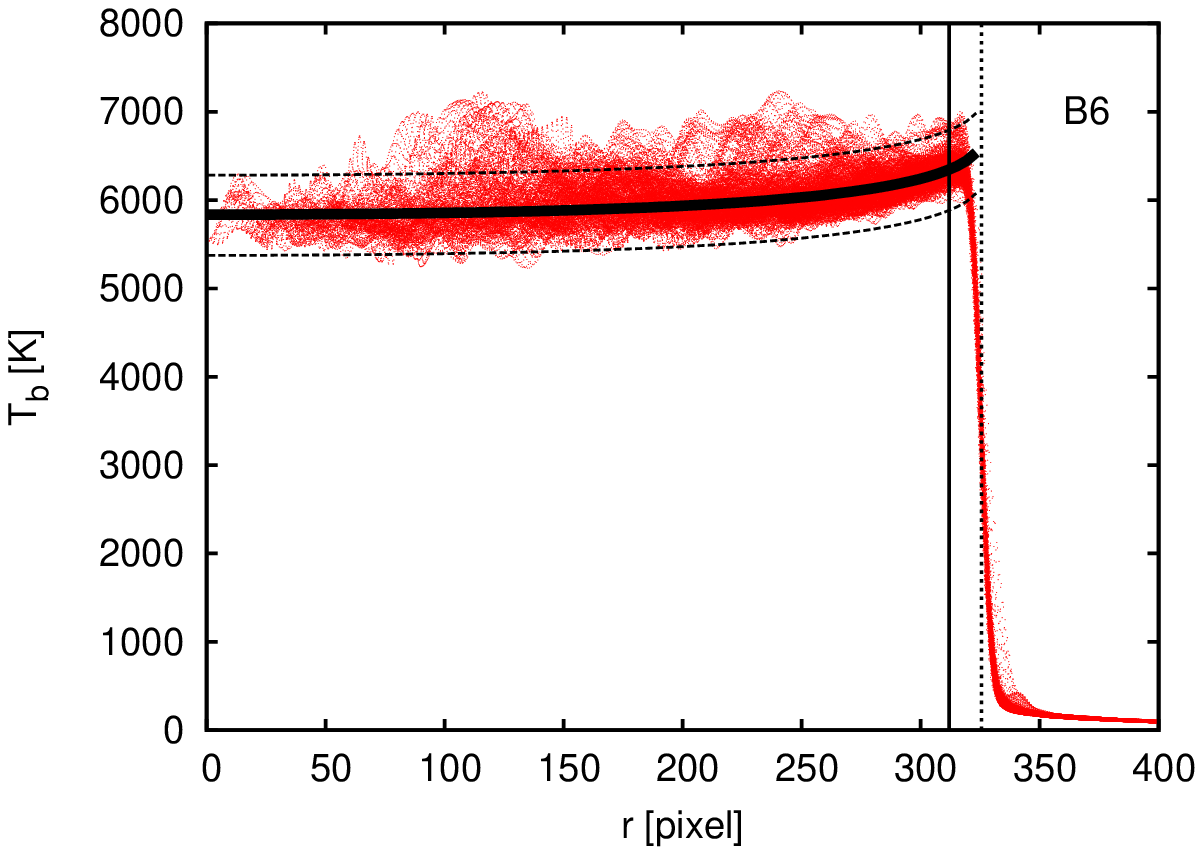}}
\caption{Brightness temperature, $T_{b}$, as a function of radial distance, $r$, from centre of the
solar disk shown with red dots. The thick solid line is the best fit function (Equation~(\ref{Eq_FitF})), while thinner dashed
lines mark the outlier boundaries above and below the best fit function. The vertical dotted line
is the solar chromospheric radius, $R_{\odot}$, and the vertical solid line is the maximum radial boundary, $r_{max}$,
used to calculate the best fit function. In the top panel we show an example in Band 3 (ID = 1, see Table~\ref{Tab_Data}) and in the bottom
panel we show an example in Band 6 (ID = 8).}
\label{Fig_CntrToLimbPlot} 
\end{figure}

Knowing the centre, we can plot $T_{b}$ as a function of pixel distance from the centre, $r$.
One such plot for each band is given in Figure~\ref{Fig_CntrToLimbPlot}. On the same image we also
show the best fit functions calculated with non-linear least-squares (NLLS) Marquardt-Levenberg algorithm
combined with automatic outlier removal based on interquartile range
with a thick solid line and outlier limits with thin dashed lines.
We see that $T_{b}$
drops suddenly before the edge as a result of the large beam size for single dish measurements
\citep{White2017}.
Therefore we fitted the data only up to an arbitrarily chosen $r_{max}$ (for Band 3 $r_{max}$=150.0 pixels
=900$^{\prime\prime}$, and for Band 6 $r_{max}$=312.0 pixels =936$^{\prime\prime}$)
which are shown as solid vertical lines in Figure~\ref{Fig_CntrToLimbPlot}.
The values of the fitted function between $r_{max}$ and $R_{\odot}$ are actually an extrapolation.

\begin{figure}
\resizebox{\hsize}{!}{\includegraphics{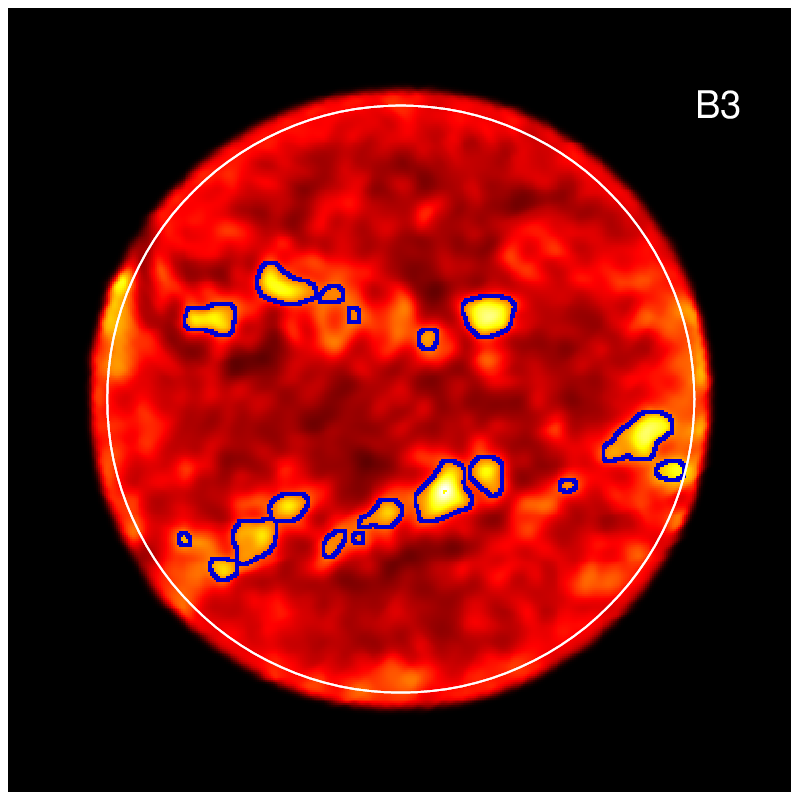}}
\resizebox{\hsize}{!}{\includegraphics{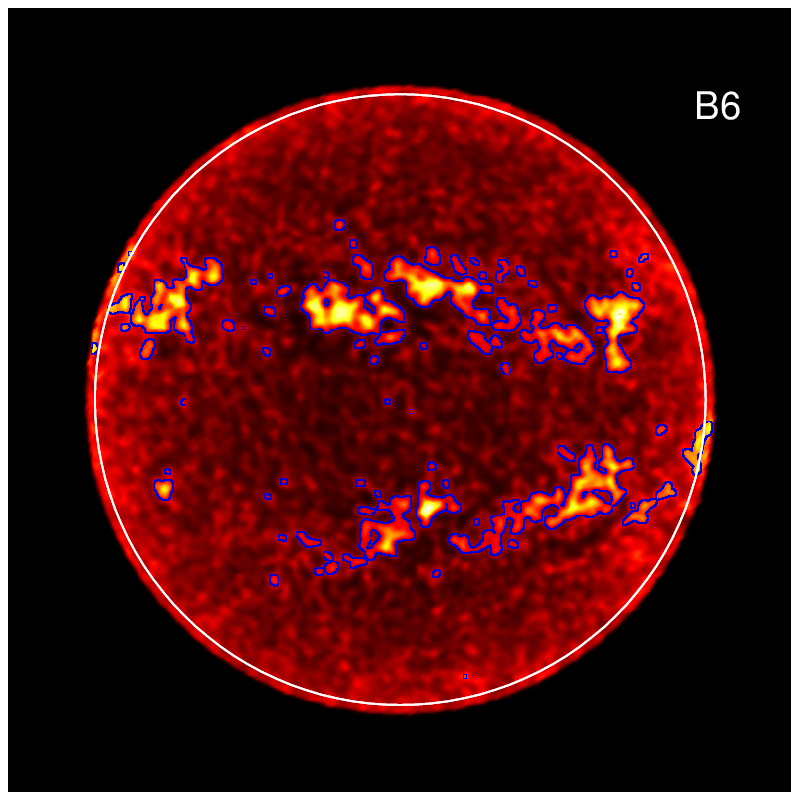}}
\caption{Regions where the outliers above the best fit function lie are marked with
thick blue line. In the top panel we show an example for Band 3 (ID = 1) and in the
bottom panel we show an example for Band 6 (ID = 8).}
\label{Fig_Outliers} 
\end{figure}
The iterative fitting procedure described in the previous Sect. marks significant amount of points
as outliers on the upper side of the fitted curve (Figure~\ref{Fig_CntrToLimbPlot}). In Figure~\ref{Fig_Outliers}
we show the position of these outliers in the full disk images. Areas where the outliers lie are
marked with a dark blue border. We can immediately conclude that these are the areas corresponding to active
regions which are significantly hotter than the quiet Sun regions \citep{Brajsa2017}. These areas
were not included in fitting of the centre to limb brightness function (Equation~(\ref{Eq_FitF})).

\begin{figure}
\resizebox{\hsize}{!}{\includegraphics{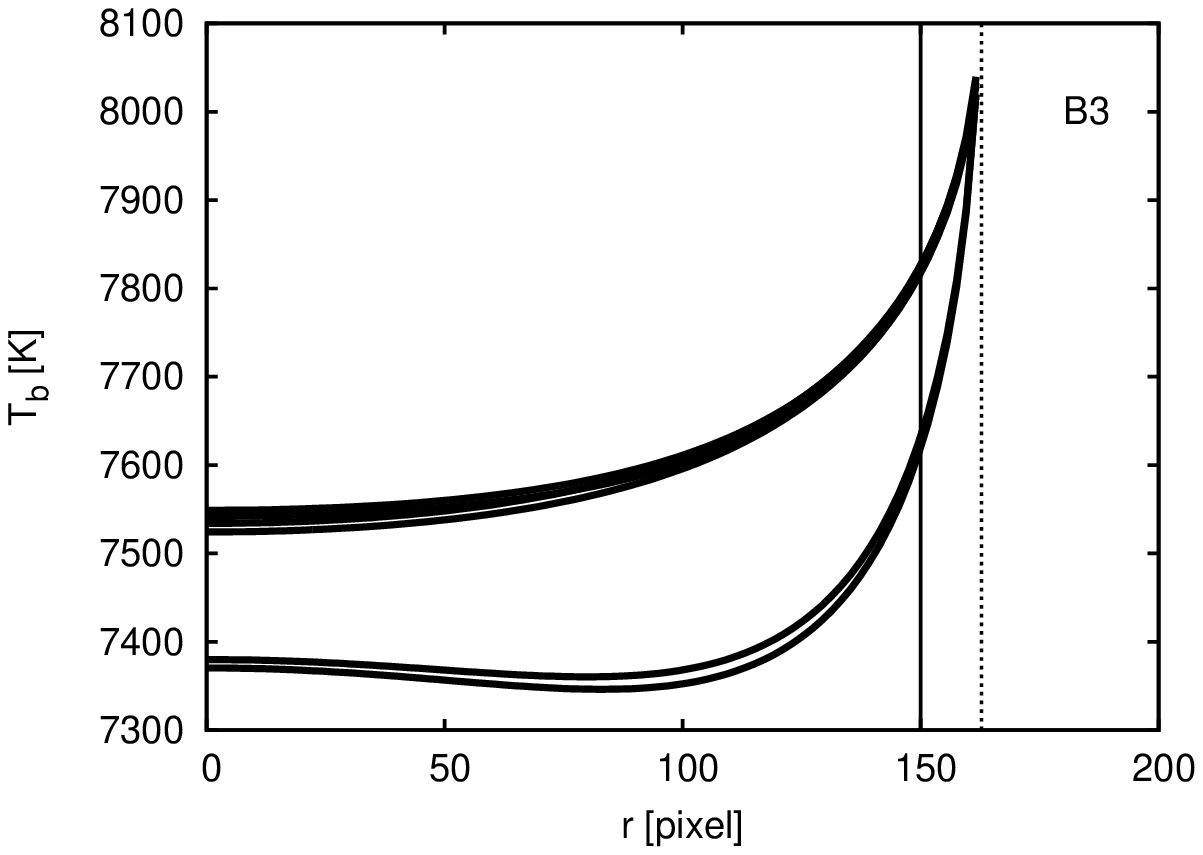}}
\resizebox{\hsize}{!}{\includegraphics{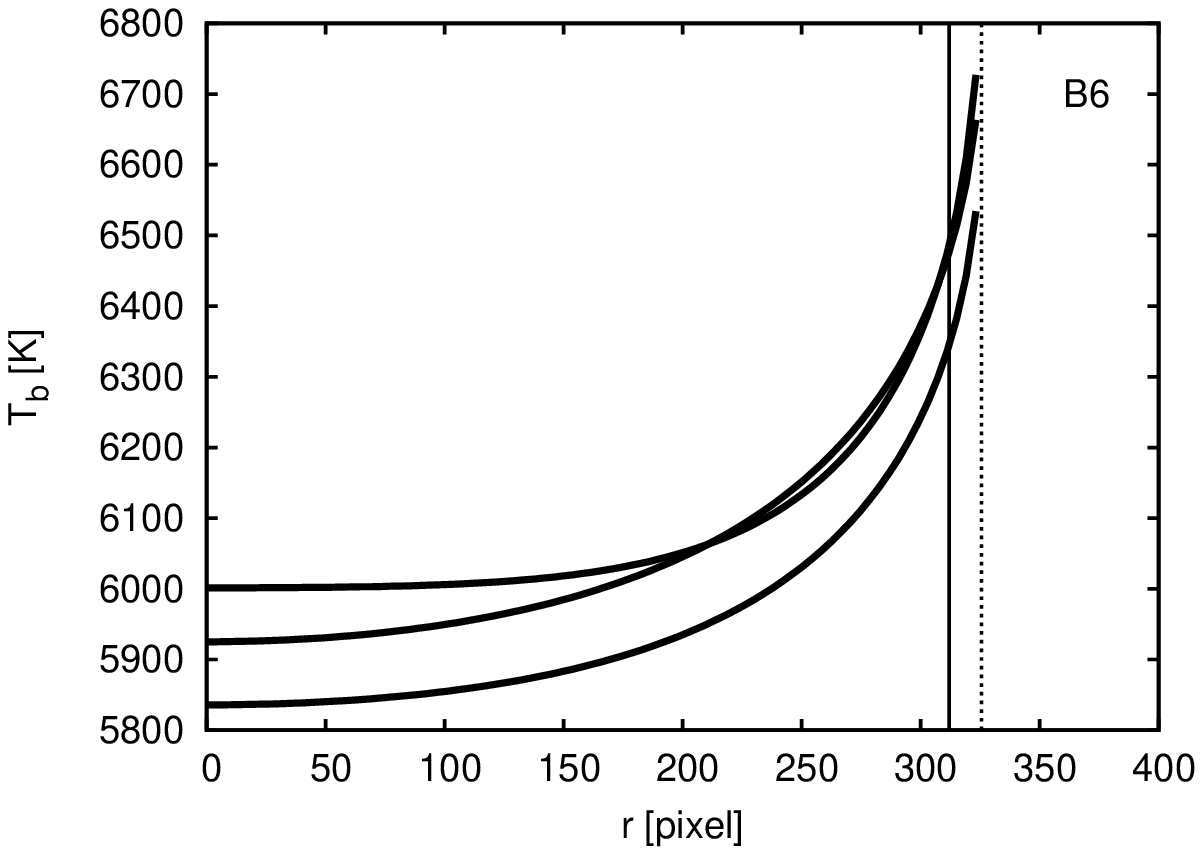}}
\caption{Best fit functions (Equation~(\ref{Eq_FitF})) for all Band 3 (top panel) and Band 6 (bottom panel) images are shown
with thick solid lines. Vertical dotted lines are chromospheric radii, $R_{\odot}$, while the thin solid vertical
lines are the values of $r_{max}$. The lowest curves in each panel correspond to the curves shown in
Fig.~\ref{Fig_CntrToLimbPlot} (IDs 1 and 8 from Table~\ref{Tab_Data}).}
\label{Fig_AllFits} 
\end{figure}

\begin{table}
\caption{Best fit coefficients with their respective errors are given in columns 3--5, the value of brightness temperature in the
centre of the solar disk which is calculated from Equation~(\ref{Eq_FitF}) by setting $\psi$= 0, is given in column 6,
and the ratio between brightness temperature at the limb and on the centre
is given in the last column.}              
\label{Tab_Coeffs_T0}      
\begin{tabular}{c c c c c c c}          
\hline
ID &Band & $A_{0}$ [K] & $A_{1}$ [K] & $A_{2}$ [K] & $T_{b}(0)$ [K] & $A_{0}/T_{b}(0)$ \\    
\hline                                   
   1 & 3 & 8256$\pm$12 & -2117$\pm$36 & 1231$\pm$25 & 7370$\pm$46 & 1.120$\pm$0.007 \\
   2 & 3 & 8248$\pm$12 & -2038$\pm$36 & 1170$\pm$25 & 7380$\pm$46 & 1.118$\pm$0.007 \\
   3 & 3 & 8141$\pm$12 &  -939$\pm$34 &  332$\pm$24 & 7534$\pm$43 & 1.081$\pm$0.006 \\
   4 & 3 & 8155$\pm$12 & -1026$\pm$34 &  413$\pm$24 & 7542$\pm$44 & 1.081$\pm$0.006 \\
   5 & 3 & 8154$\pm$12 &  -997$\pm$35 &  392$\pm$24 & 7549$\pm$44 & 1.080$\pm$0.007 \\
   6 & 3 & 8140$\pm$12 &  -966$\pm$35 &  350$\pm$24 & 7524$\pm$44 & 1.082$\pm$0.007 \\
\hline                                             
   7 & 6 & 6932$\pm$4  & -1812$\pm$12 &  881$\pm$9  & 6001$\pm$16 & 1.155$\pm$0.003 \\
   8 & 6 & 6688$\pm$4  & -1336$\pm$12 &  484$\pm$9  & 5836$\pm$15 & 1.146$\pm$0.003 \\
   9 & 6 & 6813$\pm$4  & -1291$\pm$12 &  403$\pm$9  & 5925$\pm$15 & 1.150$\pm$0.003 \\
\hline                                             
\end{tabular}
\end{table}
\begin{figure}
\resizebox{\hsize}{!}{\includegraphics{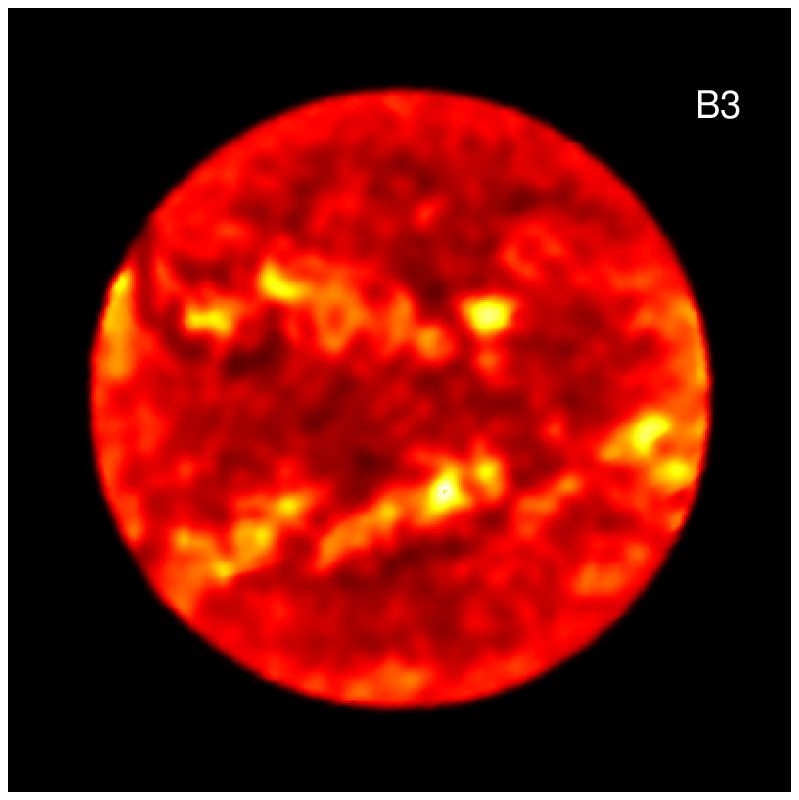}}
\resizebox{\hsize}{!}{\includegraphics{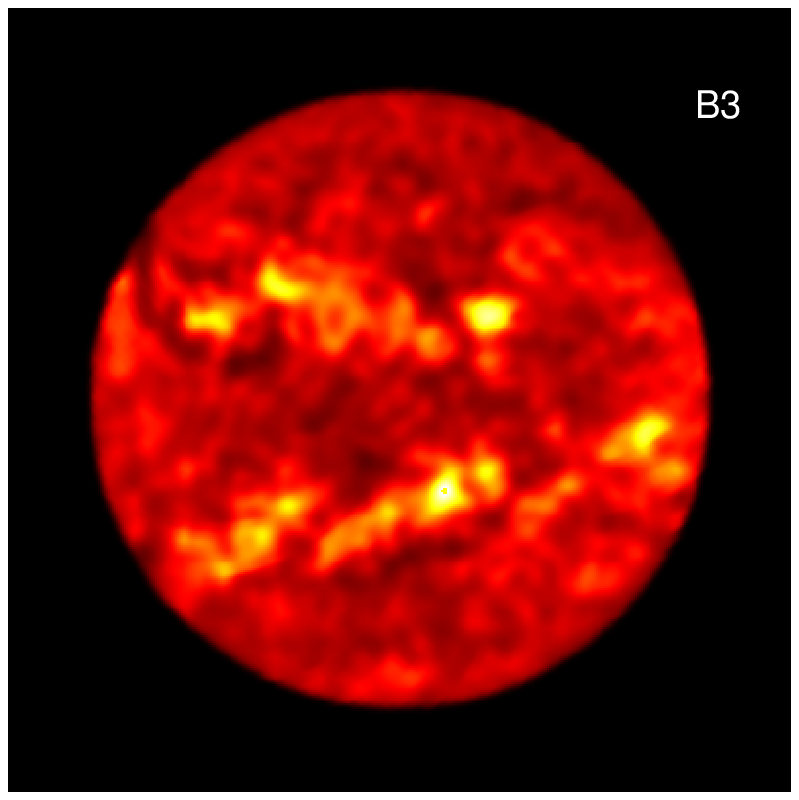}}
\caption{{\em Top panel:} Example (ID = 1) of a full disk image in Band 3 without removal of the centre to limb brightness function.
{\em Bottom panel:} The same solar disk image in Band 3 with the centre to limb brightness function (Equation~(\ref{Eq_FitF}) and Table~\ref{Tab_Coeffs_T0})
removed.}
\label{Fig_B3FlatImage} 
\end{figure}
\begin{figure}
\resizebox{\hsize}{!}{\includegraphics{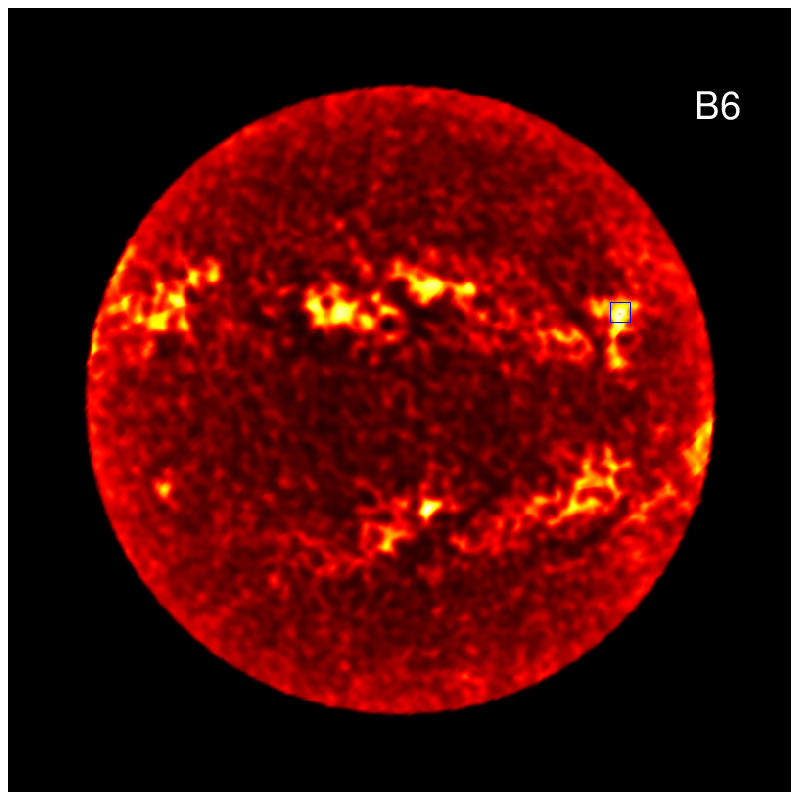}}
\resizebox{\hsize}{!}{\includegraphics{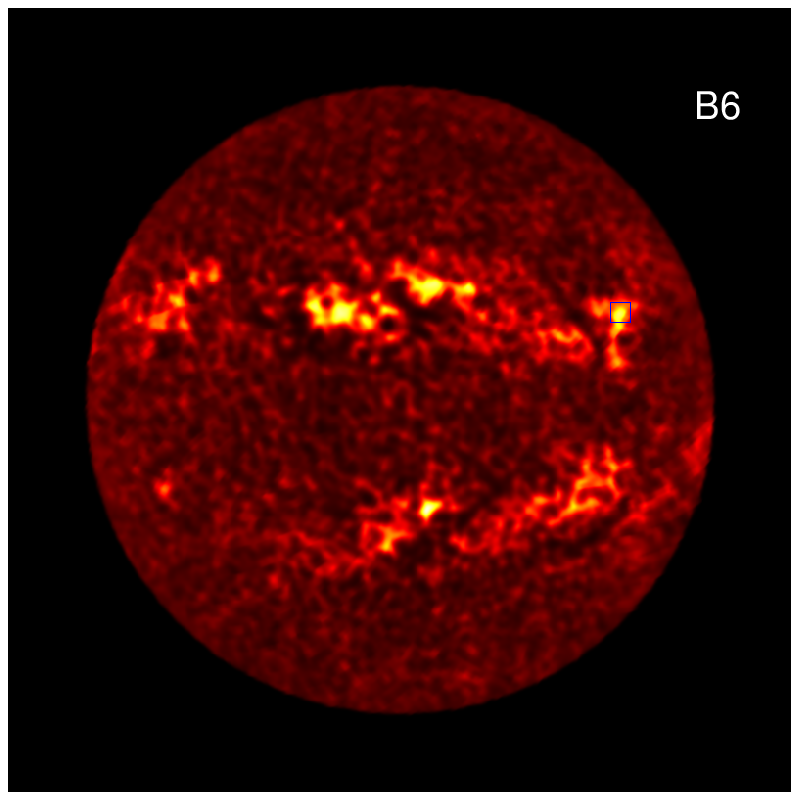}}
\caption{{\em Top panel:} Example (ID = 8) of a full disk image in Band 6 without removal of the centre to limb brightness function.
{\em Bottom panel:} The same solar disk image in Band 6 with the centre to limb brightness function (Equation~(\ref{Eq_FitF}) and Table~\ref{Tab_Coeffs_T0})
removed. Blue rectangle shows the area of one active region where we calculated average $T_{b}$ in both images for comparison.}
\label{Fig_B6FlatImage} 
\end{figure}

In Figure~\ref{Fig_AllFits} we show fits (Equation~(\ref{Eq_FitF})) for all Band 3 full disk images
in the top part of the plot and for all Band 6 images in the bottom part of the plot.
Vertical dotted lines mark the chromospheric radii, $R_{\odot}$, while the vertical solid
lines show the maximum radial boundaries, $r_{max}$, used for fitting.
The difference between the curves is about 200 K in both Bands.
The four Band 3 curves very close in value are fits to a sequence of images taken by the ALMA telescope
within half an hour (IDs from 3 to 6).

In Table~\ref{Tab_Coeffs_T0} we give values of all coefficients by using the NLLS algorithm to fit Equation~(\ref{Eq_FitF}) to each
ALMA image used in this work. In addition we also show values of brightness temperatures in the centre of the solar
disk which can be obtained by setting $\psi$=0 to Equation~(\ref{Eq_FitF}), $T_{b}(\psi=0) = A_{0} + A_{1} + A_{2}$.
The errors of the fit coefficients are also given in Table~\ref{Tab_Coeffs_T0} and
all of them are significantly smaller than the values of the coefficients suggesting that the functional
form in Equation~(\ref{Eq_FitF}) describes the data well. Similar conclusions can be drawn from the shape
of the curves compared to actual data points (Figure~\ref{Fig_CntrToLimbPlot}). The errors for Band 3, compared to
errors in Band 6, are larger. It is a consequence of fewer data points in lower resolution images.
The errors of the coefficients in Table~\ref{Tab_Coeffs_T0} describe how well is the average curve
determined from the data set. These should not be confused with the standard deviation which is larger and can be estimated
from the scatter of data points in Figure~\ref{Fig_CntrToLimbPlot}. Errors of the coefficients, which describe how
reliable is the average fit curve for the given data set, drop with the
number of data points, while the standard deviation, which describes the scatter in the distribution around
the average stabilises after sufficient number of data points is reached.

Since the fitting algorithm rejected areas where active regions are present, we suggest that the values
$T_{b}(0)$ are actually brightness temperatures of the quiet Sun in the centre of the disk.
Comparing the $T_{b}(0)$ from Table \ref{Tab_Coeffs_T0} with that of \citet{White2017}
(7300 K and 5900 K for Bands 3 and 6, respectively)
we see that the results for Band 6 are in agreement within $\approx$ 100 K while for Band 3 differences
can be up to 250 K. The differences are fairly large considering that we used the normalisation
procedure suggested by \citet{White2017}. However, in Figs.~\ref{Fig_B3FlatImage} and \ref{Fig_B6FlatImage} we can see that the full disk images show
a rich quiet Sun network which dominates the variability on small spatial scales \citep{White2017}.
Since the averaging area for normalisation is rather small (radii being $\approx$15 pixels in Band 6 and $\approx$11 pixels in Band 3)
the procedure is sensitive to chance encounter of small fairly bright or dark areas within the region.

The value of brightness temperature on the limb can be found by putting $r=R_{\odot}$ in Equation~(\ref{Eq_CosRinPix}) which
when substituted into Equation~(\ref{Eq_FitF}) gives $T_{b}(\psi = \Omega)=A_{0}$.
In Table~\ref{Tab_Coeffs_T0} we also give the ratio between the limb brightness temperature and brightness temperature
in the centre, $A_{0}/T_{b}(0)$. The ratio is smaller in Band 3 (between 8\% and 12\%) than in Band 6 ($\approx$15\%).
The values of $A_{0}$ and $A_{0}/T_{b}(0)$ are lower limits of the limb brightening effect because
of the fairly large side lobes whose effect we ignored in this study.

As a final step we produce flattened images with centre to limb brightness function removed.
In order to make such images we subtracted from each pixel the function given in Equation~(\ref{Eq_FitF}) and
Table~\ref{Tab_Coeffs_T0} and added the values of the brightness temperature in the centre of the
solar disk, $T_{b}(0)$.
The resulting images, scaled to $T_{b}(0)$, are shown in the bottom parts of Figs.~\ref{Fig_B3FlatImage}
and \ref{Fig_B6FlatImage} for Band 3 and Band 6, respectively. In the top part of Figs.~\ref{Fig_B3FlatImage}
and \ref{Fig_B6FlatImage} we provide the original non-flattened images for comparison.
The effect of removing the centre to limb brightening trend is clearly seen, especially in Band 6 with higher resolution.
These flattened images provide a way to measure $T_{b}$ of various features consistently and compare them across the disk.
For comparison we calculated the average $T_{b}$ in an area marked with the blue rectangle in top and bottom
part of Fig.~\ref{Fig_B6FlatImage}. For the top part we obtain the value $T_{b}\approx$6950 K while for the flattened
image (bottom part) the value is $T_{b}\approx$6770 K. The difference comes from the limb brightening effect which adds about
180 K at this radial distance compared to the centre of the disk.

\section{Discussion and Conclusion}

We presented a method to calculate the centre to limb brightness variations as a 2nd order polynomial function
of $\cos{\psi}$ on full disk solar images
and applied the method to 9 images made by ALMA in two spectral bands.
This approach is different than in \citet{Alissandrakis2017} and \citet{Selhorst2019} where authors calculated
the average brightness for a series of distances from the centre, $r$. In addition \citet{Alissandrakis2017} masked areas
they deemed too bright (plages) or too dark (sunspots and filament channels) and obtained a smoother looking curve
compared to \citet{Selhorst2019}. In our paper we did not mask any areas {\em a priori}, but used a procedure which
removes outliers automatically. The majority of the outliers turned out to be active regions. We can see in Table~1 of
\citet{Brajsa2017} that the difference between the active regions and local quiet Sun level
was $\approx$1000 K, while for all other features (sunspots, inversion lines, filaments, and coronal holes) the difference
was below 200 K. Therefore, the outlier removal method was capable of isolating only features significantly different
from the quiet Sun brightness temperature.
The fitting function (Equation~(\ref{Eq_FitF}))
proved to be satisfactory and the calculated coefficients are well defined when compared with their uncertainties
(Table~\ref{Tab_Coeffs_T0}).
The resulting brightness temperatures in the centre are comparable to the values of the quiet Sun $T_{b}$.
The ratio of the limb brightness temperature to the solar disk centre brightness temperature (Table~\ref{Tab_Coeffs_T0})
is larger in Band 6 ($\approx$15\%) than in Band 3 (8\% to 12\%). The values are probably underestimated
because the effect of the side lobes was ignored in this study.

\citet{Selhorst2019} investigated polar brightening in ALMA wavelengths using almost the same data as in this paper.
Our model is only dependent on $r$, not on polar coordinates $(r, \phi)$ and so it is insensitive to the polar brightening
or, in other words, dependence of $T_{b}$ on $\phi$. However, flattened images can be used to investigate possible variations
of the brightness temperature on the polar coordinate $\phi$ because the information about such variations should still
be present in the flattened images.

One advantage of using the simple function fit for limb brightening profile is that we can
easily remove the trend from the image and create the flattened images.
Flattened images can be used to compare $T_{b}$ directly between two features regardless of their distance
from the centre of the solar disk since the centre to limb variations are removed.
For example, we found in this study that an active region at radial distance $\approx$0.75 $R_{\odot}$
in Band 6
appears $\approx$ 180 K brighter than what it would be if it were located in the centre of the disk due to the
limb brightening. Comparing this to the result from \citet{Brajsa2017} for active regions ($\approx$ 1000 K above the
quiet Sun level), we can see that the effect is quite significant. The effect would be even larger for features closer to the limb.
Another potential usage might be in tracking
the evolution of $T_{b}$ for a single feature over time. Since the location, and hence its radial
distance from the solar centre of the feature changes due to the solar rotation, the
effect of limb brightening affects such observations differently in each image.
Implicit assumption here is that the brightness of a particular solar feature on various locations from the centre is
affected exactly the same as the brightness of the quiet Sun, which might not always be the case.



SV data used in this work were released with reference images taken in spectral window no. 3 (out of 4 possible
spectral windows) of both bands.
Therefore, we limited ourselves to analysis of only the spectral window no. 3 in this work.
We expect that other spectral windows,
having different base frequency, would have somewhat different centre to limb brightness profile. This is
another reason why an analysis and flattening of images similar to the one presented in this paper should
be performed on any future ALMA observation. 
Contour plots are frequently used to compare features visible at different wavelengths
\citep{Kundu1976, Brajsa1992, Gopalswamy1999, Silva2005, White2006, Brajsa2007, Loukitcheva2014, Iwai2015, Iwai2016}.
The shapes of such contours are also affected
by the centre to limb brightness variations, so flattened images may be a better choice.

The empirical method presented here is not a perfect solution. Ideally, theoretical models describing
$T_{b}$ of various features and integration of radiation transfer equation should take into account that the solar atmosphere appears
different when observed at different angles
(Figure~\ref{Fig_Geometry}) and reproduce limb brightening functions similar to the ones presented here.
This would enable to constrain electron densities as well as temperatures at various heights in the solar atmosphere.

\begin{acks}
This work has been supported by the Croatian Science Foundation under the project 7549 "Millimeter and submillimeter
observations of the solar chromosphere with ALMA". This paper makes use of the following ALMA data:
ADS/JAO.ALMA\#2011.0.00020.SV.
ALMA is a partnership of ESO (representing its member states), NSF (USA) and NINS (Japan), together with NRC (Canada),
MOST and ASIAA (Taiwan), and KASI (Republic of Korea), in cooperation with the Republic of Chile.
The Joint ALMA Observatory is operated by ESO, AUI/NRAO and NAOJ.
\end{acks} 

%

%

%

%
%

%
%
 \bibliographystyle{spr-mp-sola}
 \bibliography{cntrToLimb}  
%
%
%
%

\end{article} 
\end{document}